# Prediction of energy consumption in hotels using ANN

**Oscar Trull** [a,*], **Angel Peiró-Signes** [b] **J.Carlos García-Díaz** [a], **Marival Segarra-Oña** [b]

[a] Department Of Applied Statistics, Operational Research and Quality , Universitat Politècnica de València, Valencia, Spain

[b] Department of Business Management, Universitat Politècnica de València, Valencia, Spain

* Corresponding: otrull@eio.upv.es

**Abstract**

The increase in travelers and stays in tourist destinations is leading hotels to be aware of their ecological management and the need for efficient energy consumption. To achieve this, hotels are increasingly using digitalized systems and more frequent measurements are made of the variables that affect their management. Electricity can play a significant role, predicting electricity usage in hotels, which in turn can enhance their circularity—an approach aimed at sustainable and efficient resource use. In this study, neural networks are trained to predict electricity usage patterns in two hotels based on historical data. The results indicate that the predictions have a good accuracy level of around 2.5% in MAPE, showing the potential of using these techniques for electricity forecasting in hotels. Additionally, neural network models can use climatological data to improve predictions. By accurately forecasting energy demand, hotels can optimize their energy procurement and usage, moving energy-intensive activities to off-peak hours to reduce costs and strain on the grid, assisting in the better integration of renewable energy sources, or identifying patterns and anomalies in energy consumption, suggesting areas for efficiency improvements, among other. Hence, by optimizing the allocation of resources, reducing waste and improving efficiency these models can improve hotel's circularity.

**Keywords:**

## 1. Introduction

The tourism sector is one of the fastest growing segments in the global economy, with an estimated annual increase rate of 3.3% until 2030 (Ayuso, 2006; Campos et al., 2024). However, this sector is also energy intensive and contributes approximately 5 % of global greenhouse gas emissions (Ayuso, 2006; Casteleiro-Roca et al., 2019). Both international and local transportation, as well as comfort and entertainment in tourist facilities, are responsible for a large part of this energy consumption.

In search of a balance between economic interests and environmental concerns, the tourism sector has implemented various strategies to increase energy efficiency and reduce waste generation (Duglio et al., 2017). Hotels, in particular, are among the most energy-intensive tourism facilities, and energy consumption represents a significant portion of their operating costs (Ayuso, 2006; J. C. Eras et al., 2019a). Furthermore, in some countries, hotels are even some of the most energy-intensive buildings (J. J. C. Eras et al., 2016), and due to the expected increase in global temperatures caused by climate change, they are foresees an increase in the current energy demand of these facilities (Iturralde Carrera et al., 2023). Much of the energy consumption is focused on the production of air conditioning (Heating, Ventilation and Air Conditioning, HVAC). Through good management and with the help of IoT and data analysis, the reduction in consumption can be reduced considerably, sometimes reaching up to 50% (Pang et al., 2021).

Therefore, efficient energy use in hotels can not only improve their economic performance (Ayuso, 2006; J. C. Eras et al., 2019a; Segarra-Oña et al., 2012) but also reduce their environmental impact. Proper management of energy consumption is an opportunity to contribute to the sustainability and preservation of our planet, while ensuring a more stable and reliable operation of the hotel. To develop an efficient management and monitoring model, it is necessary to constantly make future predictions regarding energy consumption, based on the parameters available to the hotel. This is where the mathematical models developed by statisticians and forecasters come into play.

Prediction using machine learning methods is an area of knowledge that is gradually gaining more interest among the hotel sector, since with few means it is capable of providing very interesting information (Lu et al., 2022). Among the most commonly used techniques is artificial neural networks (ANN).

Casteleiro-Roca et al. (2019) propose a hybrid intelligent model for predicting energy load in hotel buildings. The model combines clustering techniques with intelligent regression methods, including Artificial Neural Networks (ANNs) and Support Vector Regression (SVR). Input parameters include energy demand, occupancy rate, and temperature. The validation using real hotel data demonstrates promising results, making this approach suitable for energy management systems in hotel resorts.

Focusing on cooling energy consumption, neural networks and support vector machines for energy prediction in a hotel building are also commonly used (Borowski & Zwolińska, 2020). Input parameters include meteorological data, time, and occupancy levels. Neural networks outperform other methods, achieving accurate predictions. The proposed model offers potential for efficient energy management in hotel buildings. In this article we propose the use of ANN to predict the energy consumption of two hotels located in Cienfuegos, Cuba. As a novelty, new climatological variables are analyzed to help improve predictions. Through this methodology, hotel managers have an additional and powerful tool to predict future energy consumption according to disparate situations, and even, with the forecast of clients and weather variables, they can determine future consumption and act accordingly.

The article is structured as follow: Section 2 explains the methodology of the conducted analysis; Section 3 shows the results obtained and Section 4 enumerates the conclusions.

## 2. Methodology

Machine learning is a field of study that, without being explicitly programmed, allows computers to learn from a given data and produce outputs that can be used for a targeted purpose. It encloses a wide range of techniques and algorithms that enable systems to recognize patterns, make predictions, and adapt based on experience. The main objective is to create intelligent systems that learn from examples and handle complex tasks (Shalev-Shwartz & Ben-David, 2014).

An Artificial Neural Network (ANN) is a mathematical and computational model inspired by the functioning of the human brain. It is designed to process information in a similar way to how neurons in the brain do(Kruse et al., 2022).

The ANN consists of a series of nodes that, intertwined with each other, allow transformations to be carried out on an input signal. This simple definition contains a more complex model, since depending on the structure, the way of organizing the nodes and the way of performing the transformations, it allows a huge number of applications with excellent results.

The nodes are called neurons, which receive information through connections with other neurons, or directly from the input variables, and perform a transformation function, generally called the activation function. Figure 1 shows the example of a neuron.

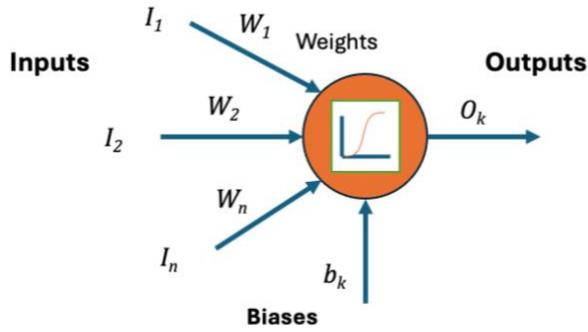

**Figure 1. Structure of an artificial neuron. Inputs and biases are processed through an activation function to produce an output.**

The formula for a single node is expressed in (1).

$$O_k = f\left(\sum_{i=1}^{n} I_i * W_i + b_k\right) \tag{1}$$

Where $O_k$ is the output as a function of the inputs $I_i$ multiplied with a weighting factor $W_i$. The function $f$, usually called activation function, could be one of the following ReLU, logistic or sigmoid, heavyside and others. In this work, we use ReLU function, defined as in (2).

$$f(v) = \begin{pmatrix} v & v \geq 0 \\ 0 & v < 0 \end{pmatrix} \tag{2}$$

Neurons are organized into organized layers so that interconnections occur between neurons. The outputs of some neurons represent the input information for other neurons, weighted as explained previously. The layers of neurons receive three different names according to the function they perform: the first are the input layer, whose information received is the input variables; the hidden layers, which relate all the information, and which is not observable, so the intermediate information is not intelligible, and that is what gives it its name; the output layer, which relates the neuron information from the hidden layers to the output. The structure is shown in Figure 2.

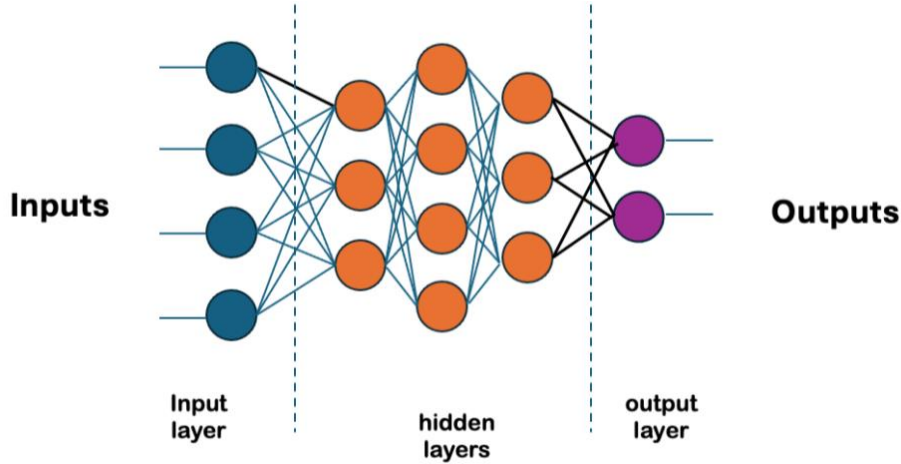

**Figure 2. Tri-layered neural network schema.**

The diagram in the figure shows a shallow neural network, with 3 layers of neurons in the hidden layer. To adjust the network and obtain the value of the weights $w_i$ and the bias $b_k$, it is adjusted iteratively, minimizing the mean square error (MSE) produced between the network output and the observed values. To improve model predictions, input data values are normalized to values between 0 and 1.

The data set used in this work belongs to a hotel located in the province of Cienfuegos, Cuba. They are provided by the article (Cabello Eras et al., 2016; J. C. Eras et al., 2019b). The data has been complemented with climatological data obtained from the Visual Crossing website (Visual Crossing Corporation, 2023). Of the global set, 90% of the data has been used for training, while the remaining 10% has been used to perform verifications. It is necessary to highlight two variables described in (Cabello Eras et al., 2016). This is the Cooling Degree Day (CDD) that reflects the influence of the outside ambient temperature $\theta_o$ compared to the reference temperature $\theta_r$, and is defined as $CDD = \sum(\theta_o - \theta_r)$. The authors also define a new variable called Room Degree Day, defined as $RDD = CDD * ORD$, where $ORD$ is the Daily Occupancy Rate.

3. Results

The energy consumed in both hotels is related to the size of the hotel, facilities and occupancy. Figure 3 shows consumption during the entire period for which data is available.

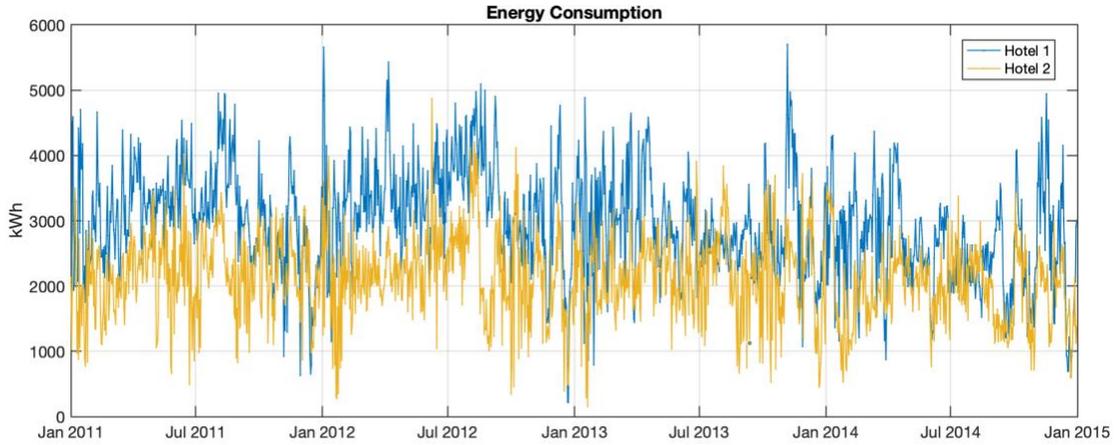

**Figure 3. Electricity consumption in both hotels 1 and 2 from 2011 until 2014**

In both cases, growth is shown during the first months of the year until the end of September, where it decreases sharply to remain at these values and repeat the cycle the following year. It is noted as a slight decrease in consumption, which is not related to the number of guests, which can be attributed to improvements in energy efficiency or weather situations.

A first exploratory analysis of the relationship between energy consumption and temperature is shown in Figure 4.

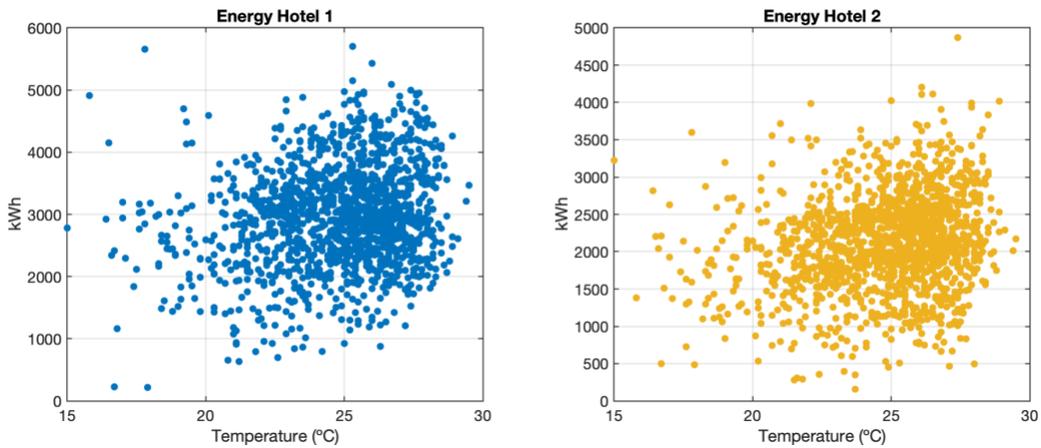

**Figure 4. Relation Energy consumed versus Air Temperature.**

It can be seen how, despite what was initially assumed, the relationship between consumption and temperature is not significant. The correlation for the case of Hotel 1 is 0.197 while for hotel 2 it is 0.217. However, as shown in Figure 5, it is observed that there is a strong relationship between consumption and RDD. The linear correlation value for Hotel 1 is 0.829 and for Hotel 2 is 0.765.

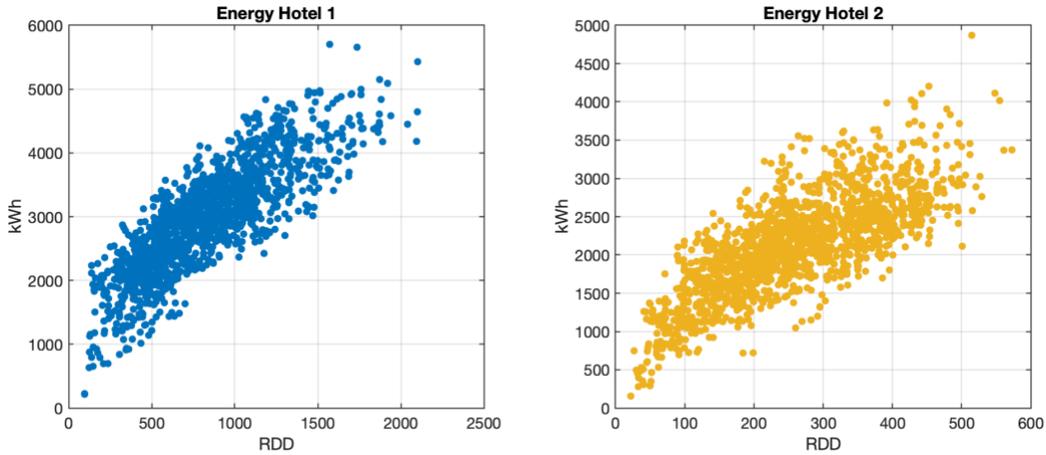

**Figure 5. Scatter plot from Energy consumption against RDD.**

The rest of the variables do not show a significant relationship with energy consumption. This is why the use of artificial intelligence models provides an advantage over traditional statistical models.

Adjustments were made to the training set to optimize the hyperparameters of the networks. The result is shown in Table 1.

**Table 1. Characteristics of the ANN and precision in the adjustment and prediction for each hotel.**

| Hotel | Neurons in layers | Activation | Fit accuracy (RMSE) | Forecast Accuracy (MAPE) |
|---|---|---|---|---|
| H1 | [230; 41; 13] | ReLU | 350.42 | 2.83% |
| H2 | [46; 32; 49] | ReLU | 571,59.56 | 2.57 % |

The same table also shows the results of the fitting error measured in terms of RMSE, as well as the prediction in terms of MAPE. Figure 6 graphically shows the results of the predictions against the actual observed data.

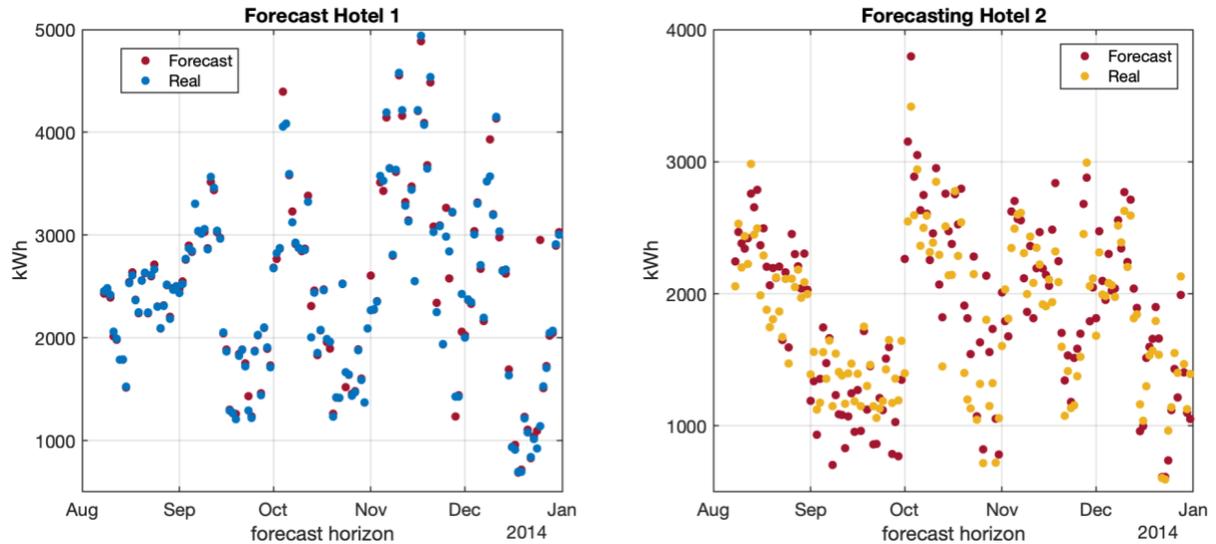

**Figure 6. Forecast comparison.**

The results obtained show that this technique is very acceptable for making predictions of energy consumption according to weather conditions and allows better planning for hotels and achieving better energy efficiency.

## 4. Conclusions

The objective of this article is to determine a methodology for predicting energy consumption in hotels through the use of machine learning tools, specifically artificial neural networks. Consumption is related to variables specific to a hotel establishment, such as the occupancy level, determined by the RDD. But there is another series of climatological variables that influence, to a lesser extent, consumption. Much of the consumption in hotels is due to HVAC. Therefore, it is important that hotels carry out correct energy management.

To carry out the study, based on the consumption data provided by (J. C. Eras et al., 2019b), an exploratory analysis has been carried out with neural network models to find those that allow accurate predictions to be made.

The results obtained by the models show a prediction accuracy of around 2.5%, which is an indicator that it is a fantastic tool for predicting consumption in hotels.